\documentclass{article}
\usepackage{amssymb}
\usepackage[dvips]{graphicx}
\begin{document}
\title{Nonequilibrium phase transitions \\ in a quasi-two-dimensional
superlattice \\ with parabolic miniband}
\author{G.M. Shmelev$^1$\thanks{E-mail: shmelev@fizmat.vspu.ru}, T.A. Gorshenina$^1$, E.M. Epshtein$^2$
 \\ \\ $^1$\emph{Volgograd State Pedagogical University, 400131, Volgograd, Russia}, \\ $^2$\emph{Institute of Radio
Engineering and Electronics, Fryazino, 141190, Russia}}
\date{}
\maketitle
\begin{abstract}
Distribution function and current density in a one-dimensional
superlattice with parabolic miniband are calculated. The current
dependence on the temperature coincides with experimental data.
Generalization is carried out to quasi-two-dimensional superlattice with
paraboloidal miniband. For a sample opened in $Y$ direction with dc
current in $X$ direction, a novel nonequilibrium phase transition is
found, namely, appearing a spontaneous transverse electric field $E_y$
under temperature rising. Near the transition temperature $T_C$,
determined by the applied field, $E_y\sim\pm\sqrt{T-T_C}$.
\end{abstract}

\section{Introduction}\label{sec1}
In present work, we study spontaneous appearance of a transverse (with
respect to the current $j_x$ in the sample) electric field $E_y$ in a
quasi-two-dimensional superlattice (2SL). In fact, a nonequilibrium
second-order phase transition (NPT2) is considered, where $E_y$ takes a
part of an order parameter, while driving field $E_x$ is a control
parameter. Such a phenomenon is similar to the multi-valued Sasaki effect
in multivalley semiconductors, that was observed in
experiments~\cite{Asche}. That effect, however, has been not treated as a
NPT2, meanwhile such an approach is fruitful enough. This approach allows
to develop fluctuation theory of NPT and reveal noise-induced NPT, as well
to investigate stochastic and vibration resonances in 2SL and other
systems~\cite{Shmelev1}--\cite{Shmelev3}. Unlike the earlier published
works~\cite{Shmelev1}--\cite{Shmelev3}, where the conventional cosine-type
model was used for the conduction miniband, here a dispersion law is
considered in form of a truncated parabola, i.~e. the dispersion law is
assumed to be parabolic up to the Brillouin zone edge. Such a problem
statement is of interest, because the modern technology allows to vary
widely the form of the potential relief and the SL energy spectrum.

In the low temperature limit ($T\rightarrow 0$) the 1SL conductivity with
parabolic miniband was studied in~\cite{Romanov}. Here we come out the
scope of the approximation used in~\cite{Romanov} and find the temperature
dependence of the current density in the model mentioned. It appears, that
other NPTs, as compared to~\cite{Shmelev1}--\cite{Shmelev3}, are possible
in 2SL with parabolic miniband. In those NPTs, the temperature is the
control parameter, so that a transverse e.m.f. appears spontaneously under
temperature raising.

\section{Distribution function and current density in 1SL}\label{sec2}
The electron energy in the 1SL lowest miniband is~\cite{Romanov}
\begin{equation}\label{1}
  \varepsilon(\mathbf p)= \varepsilon(\mathbf p_\bot)
  +\frac{p_x^2}{2m},\quad -\pi\hbar/d<p_x<\pi\hbar/d,
\end{equation}
where $\mathbf p$ is quasimomentum, $d$ is SL period, $X$ axis being
directed along the SL axis, $\varepsilon(\mathbf p_\bot)$ is in-plane
electron energy, $\pi^2\hbar^2/md^2\equiv\Delta$ is double miniband width.

In quasi-classical situation ($\Delta\gg eEd,\,\hbar/\tau$; $\tau$ being
electron momentum relaxation time, $e$ is electron charge), the current
density in uniform dc electric field $\mathbf E$ is found by solving
Boltzmann equation with collision integral within $\tau$-approximation:
\begin{equation}\label{2}
  \left(e\mathbf E,\frac{\partial F(\mathbf p)}{\partial\mathbf p}\right)
  =\frac{F_0(\mathbf p)- F_(\mathbf p)}{\tau},
\end{equation}
where $F_0(\mathbf p)$ is equilibrium electron distribution function,
$F_(\mathbf p)$ is unknown distribution function perturbed due the
electric field.

We use dimensionless variables below by changing $\pi\hbar\mathbf
p/d\rightarrow\mathbf p,\;\mathbf E/E_0\rightarrow\mathbf
E,\;T/\Delta\rightarrow T$ ($E_0=\pi\hbar/ed\tau$, $T$ is temperature in
energy units).

With the field $\mathbf E$ is directed along the 1SL axis, we have
$F(\mathbf p)=f_0(\mathbf p_\bot)f(p_x)$, $F_0(\mathbf p)=f_0(\mathbf
p_\bot)f_0(p_x)$, where $f_0(p_x)$ is equilibrium distribution function,
normalized to the carrier density $n$ ($f_0(\mathbf p_\bot)$ being
normalized to unity). Thus, $f(p_x)$ function satisfies the following
equation:
\begin{equation}\label{3}
  E_x\frac{df}{dp_x}=f_0-f\quad (-1<p_x<1).
\end{equation}

We consider nondegenerate electron gas, so that
\begin{equation}\label{4}
  f_0(p_x)=2n\left(\sqrt{2\pi T}\mathrm{erf}
  \left(\frac{1}{\sqrt{2T}}\right)\right)^{-1}\exp\left(-\frac{p_x^2}{2T}\right)
\end{equation}
($\mathrm{erf}(z)$ is error function).

In the low temperature limit ($T\rightarrow 0$), Eq.~(\ref{4}) reduces to
the function used in~\cite{Romanov}
\begin{equation}\label{5}
  g_0(p_x)=2n\delta(p_x).
\end{equation}

The exact solution of Eq.~(\ref{3}) takes the form
\begin{eqnarray}\label{6}
  &&f(p_x)=\frac{1}{E_x\mathrm{erf}\left(1/\sqrt{2T}\right)}
  \exp\left(-\frac{p_x}{E_x}+\frac{T}{2E_x^2}\right)\mathrm{erf}
  \left(\frac{p_x}{\sqrt{2T}}-\frac{\sqrt T}{\sqrt{2}E_x}\right)\nonumber
  \\  &&+C\exp\left(-\frac{p_x}{E_x}\right).
\end{eqnarray}
The second term is general solution of the homogeneous equation~(\ref{3}).
The constant $C$ is found from the function $f(p_x)$ periodicity,
$f(-1)=f(1)$. Then we get
\begin{eqnarray}\label{7}
  &&f(p_x)=\frac{n}{E_x\mathrm{erf}\bigl(1/\sqrt{2T}\bigr)}\biggl\{\exp\Bigl(-\frac{p_x}{E_x}+\frac{T}{2E_x^2}\Bigr)
  \biggl[\mathrm{erf}\Bigl(\frac{p_x}{\sqrt{2T}}-\frac{\sqrt
  T}{\sqrt{2}E_x}\Bigr)\nonumber
  \\&&-\bigl[\exp(2/E_x)-1\bigr]^{-1}\mathrm{erf}\Bigl(\frac{\sqrt
  T}{\sqrt{2}E_x}-\frac{1}{\sqrt{2T}}\Bigr)\nonumber \\ &&+\bigl[1-\exp(-2/E_x)\bigr]^{-1}\mathrm{erf}\Bigl(\frac{\sqrt
  T}{\sqrt{2}E_x}+\frac{1}{\sqrt{2T}}\Bigr)\biggr]\biggr\},\quad
  -1<p_x<1.
\end{eqnarray}

In limiting case $E_x\rightarrow 0$ Eq.~(\ref{7}) reduces to
Eq.~(\ref{4}). In another limiting case, $T\rightarrow 0$, we get the distribution function found
in~\cite{Romanov}:
\begin{equation}\label{8}
  g(p_x)=\frac{2n}{E_x}\exp\left(-\frac{p_x}{E_x}\right)\cases{[1-\exp(-2/E_x)]^{-1},
   & $0<p_x<1$,\cr
[\exp(2/E_x)-1]^{-1}, & $-1<p_x<0$}.
\end{equation}

\begin{figure}[h]
\includegraphics[scale=0.4]{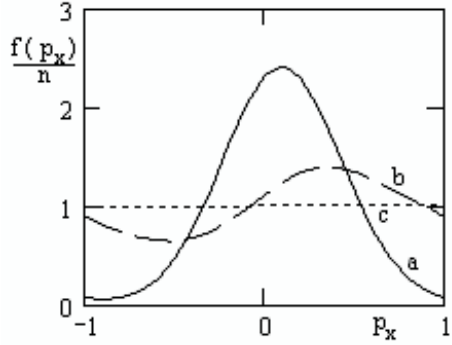}
\caption{Distribution function $f(p_x)$ at various values of the driving
field and temperature. a~--- $E=0.1,\;T=0.1$; b~--- $E=1,\;T=0.1$; c~---
$E=3,\;T=2$.}\label{fig1}
\end{figure}

The function~(\ref{7}) satisfies the same normalization condition as the
equilibrium function $f_0(p_x)$
\begin{equation}\label{9}
  \frac{1}{2}\int\limits_{-1}^1f(p_x)\,dp_x=n
\end{equation}
and, therefore, it makes the collision integral vanish. Besides, the
left-hand side of the Boltzmann equation~(\ref{3}) vanishes too, because
of the periodicity condition mentioned. The condition~(\ref{9}) can be
proved by direct calculation using formulae from~\cite{Khadzhi}. The
distribution function $f(p_x)$ at several values of $E_x$ and $T$ is shown
in Fig.~\ref{fig1}.

The current density (in dimensional units) can be found by a conventional
way:
\begin{equation}\label{10}
  j_x=\frac{ed}{2\pi\hbar m}\int\limits_{-\pi\hbar /d}^{\pi\hbar /d}p_xf(p_x)\,dp_x.
\end{equation}

\begin{figure}[h]
\includegraphics[scale=0.4]{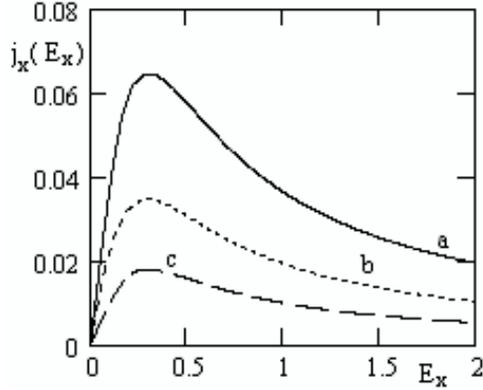}
\caption{The current density at various temperature values:
a - $T=0,5$; b - $T=1$; c - $T=2$.}\label{fig2}
\end{figure}

By substitution function~(\ref{7}) into Eq.~(\ref{10}) we get
\begin{eqnarray}\label{11}
  j_x=E_x+\bigl[2\mathrm{erf}\bigl(1/\sqrt{2T}\bigr)\sinh(1/E_x)\bigr]^{-1}
  \exp\biggl(\frac{T}{2E_x^2}\biggr)\nonumber \\
\times\biggl[\mathrm{erf}\biggl(\frac{\sqrt
T}{\sqrt{2}E_x}-\frac{1}{\sqrt{2T}}\biggr) -\mathrm{erf}\biggl(\frac{\sqrt
T}{\sqrt{2}E_x}+\frac{1}{\sqrt{2T}}\biggr)\biggr].
\end{eqnarray}
Here $j_x$ is expressed in units of $j_0=ne\Delta d/\pi\hbar$, while all
the quantities are written in dimensionless form. Equation~(\ref{11})
determines the current--voltage curve (CVC) for the parabolic miniband 1SL
with the current density temperature dependence taking into account. CVC
behavior at various temperatures is shown in Fig.~\ref{fig2} (CVC at
$T\rightarrow 0$ was shown in Fig. 1 of~\cite{Romanov}).

In limiting case $T\rightarrow 0$ Eq.~(\ref{11}) reduces to the expression
found in~\cite{Romanov}:
\begin{equation}\label{12}
  j_x=E_x-\frac{1}{\sinh(1/E_x)}.
\end{equation}

The function~(\ref{12}) reaches maximum at $E_x=0.3736818\equiv E_{C0}$.
In low fields ($|E_x|\ll E_{C0}$) we have
\begin{equation}\label{13}
  j_x=\left(1-\sqrt{\frac{2}{\pi T}}\frac{\exp(-1/2T)}{\mathrm{erf}(1/\sqrt{2T})}\right)E_x
  =\left\langle\frac{p_x^2}{T}\right\rangle_0E_x,
\end{equation}
where angle brackets mean averaging over the equilibrium distribution.
Note that the mobility temperature dependence in low fields (the
expression within round brackets in Eq.~(\ref{13})) is close to the
analogous dependence for the miniband cosine model
($I_1(1/2T)/I_0(1/2T)$, $I_n(z)$ being the modified Bessel function).

Note, that the current maximum $E_C(T)$ shifts leftwards (at $E_x>0$) and
tends to its limiting value $E_1=0.2930031$ with rising temperature.
Essentially, the $E_C$ value does not depend on the temperature at all in
the cosine model. Meanwhile, in experiment~\cite{Grahn} a shift of $E_C$
to lower fields has been found with raising temperature. (In~\cite{Grahn}
miniband transport in 1SL GaAs/AlAs was investigated in temperature range
10--300 K under fields up to 1.5 kV/cm, and negative differential
conductivity was found surely enough.)

\begin{figure}[h]
\includegraphics[scale=0.4]{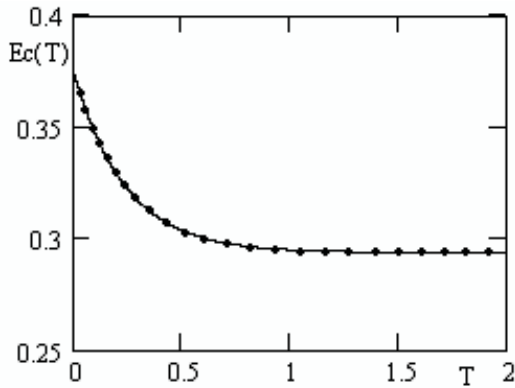}
\caption{Temperature dependence of the current maximum (Eq.~(\ref{11})).
The dotted line shows approximation~(\ref{14}).}\label{fig3}
\end{figure}

We investigated numerically the temperature dependence $E_C(T)$. That
function can be approximated (with precision in calculations of $10^{-4}$)
with the following formula:
\begin{equation}\label{14}
  E_C(T)=\sqrt{E_1^2+(E_{C0}^2-E_1^2)\exp(-\beta T)}\quad (E_x=\mathrm{const},\;E_{C0}=E(0)),
\end{equation}
where $\beta=4.2$. In Fig.~\ref{fig3}, the numerical solution ($E_C(T)$)
of the equation $\partial j_x/\partial E_x=0$ is shown, the dotted curve
being the approximation~(\ref{14}).

\section{Nonequilibrium phase transitions in 2SL}\label{sec3}
The dispersion law of the square 2SL mentioned in Section~\ref{sec1} takes
the form
\begin{equation}\label{15}
  \varepsilon(\mathbf p)=\varepsilon_0+\frac{p_x^2+p_y^2}{2m},\quad -\pi\hbar/d<p_{x,\,y}<\pi\hbar/d,
\end{equation}
where $\varepsilon_0=\mathrm{const}$ is electron energy in the lowest
size-quantized level, $\Delta$ is miniband width.

If the coordinate axes are directed along the SL principal ones, then the
current density along $X$ axis is given by Eq.~(\ref{11}), while that
along $Y$ axis is given by the same formula with $x\rightarrow y$ change.

Now a situation is considered, where the coordinate axes are rotated by
angle $45^\circ$ with respect the principal axes of the square 2SL. In
this case we have
\begin{eqnarray}\label{16}
  j_y=E_y+\left[4\mathrm{erf}\left(1/\sqrt{2T}\right)\sinh\left(1/(E_x+E_y)\right)\right]^{-1}\exp\left(\frac{T}{2(E_x+E_y)^2}\right)\nonumber \\
\times\left[\mathrm{erf}\left(\frac{\sqrt
T}{\sqrt{2}(E_x+E_y)}-\frac{1}{\sqrt{2T}}\right)
-\mathrm{erf}\left(\frac{\sqrt T}{\sqrt{2}(E_x+E_y)}+\frac{1}{\sqrt{2T}}\right)\right]\nonumber \\
-\left[4\mathrm{erf}\left(1/\sqrt{2T}\right)\sinh\left(1/(E_x-E_y)\right)\right]^{-1}\exp\left(\frac{T}{2(E_x-E_y)^2}\right)\nonumber \\
\times\left[\mathrm{erf}\left(\frac{\sqrt
T}{\sqrt{2}(E_x-E_y)}-\frac{1}{\sqrt{2T}}\right)
-\mathrm{erf}\left(\frac{\sqrt
T}{\sqrt{2}(E_x-E_y)}+\frac{1}{\sqrt{2T}}\right)\right],
\end{eqnarray}
where the field is expressed in units of $\sqrt{2}\pi\hbar/ed\tau$, while
the current in units of $\sqrt{2}\pi\hbar/ed\tau$. The formula for $j_x$
is obtained from Eq.~(\ref{16}) by interchange $y\leftrightarrow x$.

Let the sample is open in $Y$ direction, so that
\begin{equation}\label{17}
  j_y=0.
\end{equation}
Then, by substituting Eq.~(\ref{16}) into Eq.~(\ref{17}), we obtain an
equation to find the spontaneous transverse field $E_y$. At first,
consider the case of extremely low temperatures ($T\rightarrow 0$). In
that case we get from Eq.~(\ref{16})
\begin{equation}\label{18}
  j_y=E_y-\frac{1}{2}\left[\frac{1}{\sinh(1/(E_x+E_y))}- \frac{1}{\sinh(1/(E_x-E_y))}\right].
\end{equation}

The solution of Eq.~(\ref{17}) in reference to the current~(\ref{18}) has
the form
\begin{equation}\label{19}
  E_{ys}=\cases{0, & $0\le |E_x|<E_{C0}$, \cr
 \pm\sqrt{\frac{\cosh(1/E_x)}{\sinh^2(1/E_x)}-E_{C0}^2},
   & $|E_x|\ge E_{C0}$}.
\end{equation}

We write down, for comparison, the analogous formula for the 2SL cosine
model:
\begin{equation}\label{20}
  E_{ys}=\cases{0, & $0\le |E_x|<1/\pi$, \cr \pm\sqrt{E_x^2-1/\pi^2}, & $|E_x|\ge 1/\pi$}.
\end{equation}

\begin{figure}[h]
\includegraphics[scale=0.4]{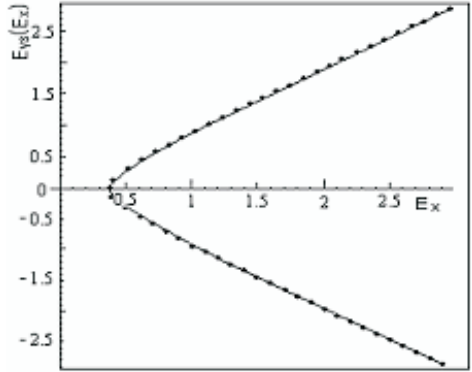}
\caption{The transverse field $E_{ys}$. The solid line shows numerical
solution of Eqs.~(\ref{17}),~(\ref{18}); the dotted one shows
function~(\ref{19}).}\label{fig4}
\end{figure}

The numerical solution of Eq.~(\ref{17}) and function~(\ref{19}) (the
dotted curve) are shown in Fig.~\ref{fig4}.

\begin{figure}[h]
\includegraphics[scale=0.35]{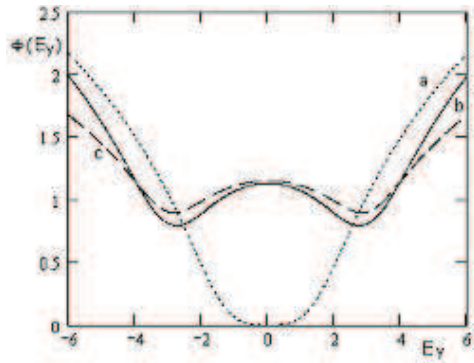}
\caption{Synergetic potential (up to a constant): a - $E_x=E_{C0}$; b -
$E_x=1$; c - $E_x=1$ for the cosine dispersion law.}\label{fig5}
\end{figure}

The solution stability against small enough fluctuations of $E_y$ field is
determined with the following inequality~\cite{BonchBruevich}:
\begin{equation}\label{21}
  \frac{\partial j_y}{\partial E_y}=\frac{\partial^2\Phi}{\partial
  E_y^2}>0,
\end{equation}
where $\Phi=\int j_ydE_y+\mathrm{const}$ is a synergetic potential (the
entropy production~\cite{Shmelev4}). Conditions~(\ref{17}) and~(\ref{21})
are satisfied at $E_y=E_{ys}$. Therefore, the solutions~(\ref{19})
and~(\ref{20}) correspond to minima of the potential $\Phi$, which becomes
double-well one (Fig.~\ref{fig5}). Thus, there is a NPT2 here.

At $T\ne 0$, the transverse field $E_y$ behaves the same way as at
$T\rightarrow 0$ with the only difference that the bifurcation point $E_C$
shifts leftwards with rising $T$ (for $E_x>0$), in complete correspondence
with behavior of the current $j_x$ maximum (see Fig.~(\ref{fig3})).

It follows from the temperature dependence of $E_C(T)$, that a novel type
NPT2 is possible in the model considered. Indeed, in the field interval
$0.317<|E_x|<E_{C0}$ the transverse field $E_{ys}(T)$ obeys the following
relationship:
\begin{equation}\label{22}
  E_{ys}(T)=\pm\sqrt{\left(E_{C0}^2-E_1^2\right)\left[\exp\left(-\beta
  T_C\right)- \exp\left(-\beta T\right)\right]}\quad (E_x=\mathrm{const}),
\end{equation}
where the critical temperature is
\begin{equation}\label{23}
  T_C=\frac{1}{\beta}\ln\left(\frac{E_{C0}^{2}-E_1^2}{\cosh(1/E_x)/\sinh^2(1/E_x)-E_1^2}\right).
\end{equation}

\begin{figure}[h]
\includegraphics[scale=0.3]{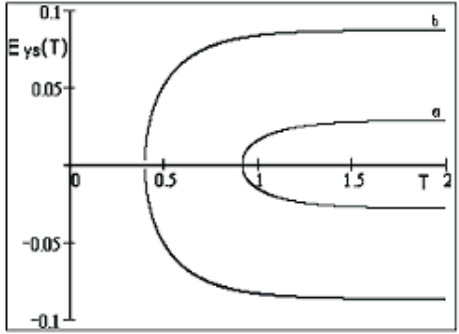}
\caption{The transverse field $E_{ys}$ at various values of the driving
field: a - $E_x=0.318$; b - $E_x=0.330$.}\label{fig6}
\end{figure}

Near the transition point $T_C$, the transverse field behaves as
$E_{ys}\sim\pm\sqrt{T_C-T}$. The NPT2 is illustrated in Fig.~\ref{fig6}.

\section{Conclusion}\label{sec4}
In present paper, an exact distribution function has been found of the
carriers in the lowest parabolic miniband of 1SL. The novel formula for
the current density in 1SL contains temperature dependence, which leads to
the current maximum shift to the low field side with rising temperature,
that agrees with available experimental data.

The considered SL model includes NPT's studied earlier. Besides, a novel
type NPT has been found, in which the sample temperature plays a role of a
control parameter.

The estimates of the effects predicted are reduced, in general, to
estimate of electric field required. At $d=10^{-7}\,\rm{cm}$,
$\tau=10^{-12}\,\rm{c}$, we have $E_0\approx 2\times10^3\,\rm{V/cm}$. At
$E_x\approx 600\,\rm{V/cm}$,  the value $T_C\approx 45\,\rm K$ is
obtained.

\end{document}